\documentclass[conference]{IEEEtran}
\usepackage{graphicx}
\usepackage{algorithm}
\usepackage{algorithmic}
\usepackage{multirow}
\usepackage{amsmath}
\usepackage{xcolor}
\usepackage{subfig}
\usepackage[top=0.78in, bottom=0.78in, left=0.72in, right=0.72in]{geometry}

\ifCLASSINFOpdf
\else
\fi
\begin{document}
\title{Distributed Edge Caching Scheme Considering the Tradeoff Between the Diversity and Redundancy of Cached Content}
\author{\IEEEauthorblockN{Shuo Wang, Xing Zhang, Kun Yang, Lin Wang, and Wenbo Wang}
\IEEEauthorblockA{Wireless Signal Processing and Network Laboratory\\
Beijing University of Posts and Telecommunications,\\ Beijing, 100876, P.R. China\\
Email: wangsh@bupt.edu.cn, hszhang@bupt.edu.cn}
\vspace*{-20pt}
}


\maketitle

\begin{abstract}
Caching popular contents at the edge of cellular networks has been proposed to reduce the load, and hence the cost of backhaul links. It is significant to decide which files should be cached and where to cache them. In this paper, we propose a distributed caching scheme considering the tradeoff between the diversity and redundancy of base stations' cached contents. Whether it is better to cache the same or different contents in different base stations? To find out this, we formulate an optimal redundancy caching problem. Our goal is to minimize the total transmission cost of the network, including cost within the radio access network (RAN) and cost incurred by transmission to the core network via backhaul links. The optimal redundancy ratio under given system configuration is obtained with adapted particle swarm optimization (PSO) algorithm. We analyze the impact of important system parameters through Monte-Carlo simulation. Results show that the optimal redundancy ratio is mainly influenced by two parameters, which are the backhaul to RAN unit cost ratio and the steepness of file popularity distribution. The total cost can be reduced by up to 54\% at given unit cost ratio of backhaul to RAN when the optimal redundancy ratio is selected. Under typical file request pattern, the reduction amount can be up to 57\%.
\end{abstract}


%
\IEEEpeerreviewmaketitle

\section{Introduction}
With the rapid growth of traffic demands in future fifth generation (5G) celular networks, various new technologies are studied to accommodate these challenges. Local caching at the edge of network (base stations and mobile devices) is one of the disruptive technologies \cite{Five}. Edge caching can reduce network load, especially backhaul traffic, by storing the most frequently requested contents at local caches. It is an approach to strike a balance between data storage and data transfer. Caching is more effective in today's information-centric network \cite{info}. As the capacity of radio access network (RAN) increases because of advances in technology, the capacity challenge and congestion problem are shifted to backhaul links connecting core network (CN) and RAN \cite{Video}. Because of the difference of users' interest, the popularity of network files differ from each other. A small number of files receive a large portion of user requests. Therefore, caching these files can satisfy most of the user demands.

Based on the all-IP architecture of current cellular networks, caches can be deployed in the CN and RAN \cite{air}. Many researches have been conducted to find the best caching policies. In \cite{Femto}, a distributed caching problem is formalized where mobile devices have access to multiple caches. Then the authors solve this NP-hard problem with approximation algorithms. They demonstrate that distributed caching can alleviate the bottlenecks in wireless video delivery. Authors in \cite{Rout} consider the bandwidth constraint of base stations (BS), and jointly optimize the caching and routing scheme to increase hit ratio of small cell BSs. The caching policy in \cite{multi} takes into consideration of multicast transmission to achieve lower traffic compared to unicast scheme. Cache replacement strategy has also been investigated intensively such as recency-based, frequency based and distributed strategies \cite{recency} \cite{freq} \cite{repl}. In \cite{mod}, the coupling of caching problem with physical layer performance is  discussed, giving out the outage probability and average delivery rate of cache-enabled small cell networks.

In this paper, we investigate the tradeoff between the diversity and redundancy of contents cached in BSs. In order to minimize the backhaul traffic between CN and RAN, the contents should be cached with the highest diversity. That is, only one copy of any content is cached in the RAN so that the most number of different contents can be cached in the RAN with limited storage. However, this will increase the transmission cost among BSs. To minimize the transmission cost among BSs, the same most popular contents should be cached at each BS, so most requests are served locally without transmission among BSs. We try to find the optimal configuration between caching diversity and redundancy to minimize the total transmission cost in the network.

First, we discuss the impact of edge caching on backhaul transmission and RAN transmission cost respectively. Then a popularity-based caching policy is proposed considering the redundant storage of most popular contents among BSs. Accordingly, an optimal redundancy caching problem (ORCP) is formulated for obtaining the optimal redundancy ratio under given system configuration. Since this is a discrete variable optimization problem, we transform it into a continuous form and obtain the optimal value of redundancy ratio with particle swarm optimization (PSO) algorithm. Finally, we analyze the factors that influence the optimal value of redundancy ratio.

The main contributions of this paper are summarized as follows:
\begin{enumerate}
\item The optimal redundancy caching problem (ORCP) is fomulated. We propose a caching scheme considering the tradeoff between caching diversity and redundancy.
\item We solve the problem with a heuristic algorithm, the particle swarm optimization algorithm. The parameters are adapted to improve the accuracy of the result.
\item We evaluate the theoretic and simulation results of total transmission cost of different redundancy ratios. We analyze the main parameters that influence the transmission cost and the value of optimal redundancy ratio.
\end{enumerate}

The rest of the paper is organized as follows. In Section~\ref{sec:sys}, the system model is described and the ORCP problem is formulated. In Section~\ref{sec:opt}, the calculation of transmission cost is presented and the problem is solved with adapted PSO algorithm. Section~\ref{sec:sim} presents our simulation results and a detailed analysis of system parameters. Finally, conclusions are drawn and future works are discussed in Section~\ref{sec:conc}.
\section{System Model and Problem Formulation}
\label{sec:sys}
In this section, the network model of mobile edge caching is firstly introduced and a popularity based caching scheme is presented. Then the edge caching redundancy ratio optimization problem is formulated.
\subsection{System Model}
\begin{figure}[!t]
\centering
\includegraphics[width= 7cm]{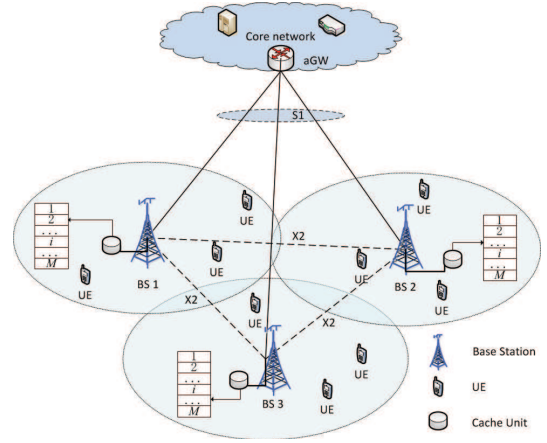}
\caption{Illustration of the network model. The caching-enabled BSs are connected with each other via X2 link.}
\label{fig_1}
\vspace*{-10pt}
\end{figure}

We consider the scenario of a cellular network in which the base stations(BS) are caching enabled as depicted in Fig.~\ref{fig_1}. The network consists of a core network and a set of $N=|\mathcal{N}|$ BSs, serving the requests of users in their coverage areas. Poisson point process (PPP) is used to model the the distribution of the BSs in a circular area of radius $r$. The density of BSs in this area is denoted as $ \lambda $. Each BS $n \in \mathcal{N}$ has a cache of size $M>0$ bytes. Let $\mathcal{F}$ denote the set of all the different files in the system so that the total number of different files in the network is $F=|\mathcal{F}| \ge MN$. The popularity of the files follows Zipf distribution with exponent $s$. For convenience, we assume that each file has the same file length normalized to 1 byte. This is reasonable because files of different length can be divided into groups of the same length. Thus, the maximum number of files each BS can store is $M$.

The BSs are connected with each other so they can share their cached files. They are connected to the core network via backhaul links. The most popular files in the network are cached in BSs and the requests of these files are served directly from the cache of each BS. If a user of BS $i$ requests for a file which is stored not in BS $i$ but in BS $j$, BS $i$ will ask BS $j$ to transfer this file to it and then serve the request of this user. We denote with $\alpha \ge 0$ as the transmission cost per byte (in monetary units/byte) among BSs, which is called unit RAN transmission cost. The requests of files which are not cached in BSs will be served by fetching files from the core network via backhaul links. Let $\beta \ge \alpha$ denote the transmission cost per byte from BS to the core network, which is called unit backhaul transmission cost. Let $\mu_{BR}=\alpha / \beta$ denote the unit cost ratio between backhaul transmission and RAN transmission.

\subsection{Problem Formulation}
As is known that caching popular contents at the edge of the network can offload backhaul traffic, but will increase traffic among BSs, so what is the optimal number of different files that should be cached in the RAN in order to minimize total transmission cost in the system? Different BSs can store the same most popular files, or they can store files that are different from others. Let us consider two extreme situations. On the one hand, when the BSs all store the same most popular files in the system, the transmission cost in the RAN will be 0 because no file is transferred among BSs. But in this way, the total number of different files cached in the RAN is the least, so the backhaul transmission cost is the highest. On the other hand, when the BSs store files that are different from each other, the total number of different files cached in the RAN is the most. In this situation, the backhaul transmission cost is the lowest since most of the file requests are served in the RAN. However, the RAN transmission cost will be the highest because of file transferring among BSs. This is referred to as the exploration vs. exploitation paradigm \cite{Living}. In order to strike a good balance between cached and uncached contents, we propose a caching scheme considering file redundancy ratio to find the optimal redundancy ratio that minimizes the total transmission cost in the system.

All the $F$ files in the system are sorted according to their popularity rank. The first $R$ files cached in all the BSs are the most popular $R$ files in the system. These files are defined as redundant files. The remaining $M-R$ files cached in each BS are different from others and are stored according to their popularity ranks as shown in Fig.~\ref{fig_2}. These files are defined as BS-specific files. We denote with $\eta=R/M$ the redundancy ratio of cached contents in BSs. According to Fig.~\ref{fig_2}, we can obtain that the popularity ranks of the cached files except for the first $R$ files are as follows:
\begin{figure}[!t]
\centering
\includegraphics[width= 8cm]{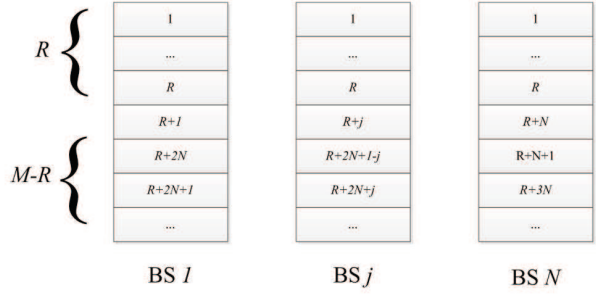}
\caption{Illustration of the caching scheme. The number in each cache unit is the popularity rank of the file stored in it.}
\label{fig_2}
\vspace*{-10pt}
\end{figure}

\begin{equation}
\label{k}
k = \left\{ {\begin{array}{*{20}{c}}
{R + (m - 1)N + j}&{m = 1,3,5,...}\\
{R + mN + 1 - j}&{m = 2,4,6,...}
\end{array}} \right.
\end{equation}
where $m = 1,2,3,...,M - R$, is the number of storage units caching BS-specific contents. Since the popularity of files follows Zipf distribution \cite{zip}, the probability that the file of rank $k$ is requested is
\begin{equation}
\label{pop}
f\left( {k,s,F} \right) = \frac{{1/{k^s}}}{{\mathop \sum \nolimits_{n = 1}^F \left( {1/{n^s}} \right)}}
\end{equation}
we denote $f_j$ as the probability that the BS-specific files cached in BS $j$ are requested, and $f_{Bh}$ as the probability that the files not cached in RAN are requested. Then we can calculate the transmission cost in RAN $c_{RAN}$ and the transmission cost of backhaul links $c_{Bh}$ as follows.
\begin{equation}
\label{cran}
{c_{RAN}} = \alpha \sum\limits_{i = 1}^N {\sum\limits_{j = 1}^N } {f_j}
\end{equation}
\vspace*{-10pt}
\begin{equation}
\label{cbh}
{c_{Bh}} = \beta {f_{Bh}} = \alpha \mu_{BR} {f_{Bh}}
\end{equation}

Therefore, the total transmission in the network is $c_{total}={c_{RAN}}+{c_{Bh}}$. Our goal is to minimize the total transmission cost by adjusting the redundancy ratio of cached contents in BSs. The redundancy ratio optimization problem is formulated formally as
\begin{equation}
\mathop {\min }\limits_\eta \qquad {c_{Bh}} + {c_{RAN}}\\
\end{equation}
\vspace*{-10pt}
\begin{equation}
\label{eqnr1}
s.t. \quad\qquad \eta  \in [0,1]\\
\end{equation}
\vspace*{-10pt}
\begin{equation}
\qquad \qquad \eta M \in {\rm{\mathcal{Z}}}
\label{eqn}
\end{equation}
where $\mathcal{Z}$ is the set of integers. (\ref{eqnr1}) indicates the range of caching redundancy ratio. (\ref{eqn}) indicates the discrete nature of the optimization variable. We call the above problem the \emph{Optimal Redundancy Caching Problem} (ORCP).
\section{Redundancy Ratio Optimization with PSO Algorithm}
\label{sec:opt}
In this section, we discuss the optimal solution of the above ORCP problem. Firstly, we will calculate the transmission cost in RAN and backhaul links respectively. Then we will solve this discrete optimization problem with PSO algorithm.

\subsection{Transmission Cost Calculation}

\subsubsection{RAN Transmission cost}
In equation (\ref{cran}), in order to obtain RAN transmission cost, we need to derive the calculation of $f_j$. $f_j$ can be calculated through the sum of the probability of each BS-specific file which BS $j$ stores being requested.

According to equation (\ref{k}) and (\ref{pop}), we can derive the value of $f_j$ as follows:

when $M-R>0$ and is even:
\begin{eqnarray}
\label{eqn_s}
{f_j} = \sum\limits_{t = 1}^{(M - R)/2} {[f(R + (2t - 2)N + j,s,F) + } \nonumber\\
f(R + 2tN + 1 - j,s,F)]
\end{eqnarray}

when $M-R>1$ and is odd:

\begin{eqnarray}
{f_j} = \sum\limits_{t = 1}^{(M - R - 1)/2} {[f(R + (2t - 2)N + j,s,F) + }\nonumber\\
 f(R + 2tN + 1 - j,s,F)] + \nonumber\\
 f(R + (M - R - 1)N + j,s,F)
\end{eqnarray}

when $M-R=1$:
\begin{equation}
{f_j} = f(R + j,s,F)
\end{equation}

Then we can obtain the RAN transmission cost based on equation (\ref{cran}).

\subsubsection{Backhaul Transmission Cost}
In equation (\ref{cbh}), $f_{Bh}$ is needed to be calculated in order to attain the value of backhaul transmission cost. The total number of different files cached in the RAN is $R + (M - R)$, the remaining files are fetched from the core network via backhaul links. The probability of these files being requested $f_{Bh}$ is:
\begin{equation}
\label{eqn_e}
f_{Bh} = \sum\limits_{k = R + (M - R)N + 1}^F {f(k,s,F)}
\end{equation}
Then the backhaul transmission cost can be obtained based on equation (\ref{cbh}).

\subsection{Optimal Solution with PSO Algorithm}
As is discussed in section~\ref{sec:sys}, the ORCP problem is discrete. In order to solve this problem, we will transform it in to a continuous problem, which is equivalent to the original problem. Since $\eta=R/M$, where $R$ and $M$ are both integers, so the value of $\eta$ is discrete. We can let $\eta$ be continuous by taking the floor of $R$ calculated from $\eta$, i.e., $R = \left\lfloor {\eta M} \right\rfloor $. Then we substitute the $R$ in equations from (\ref{eqn_s}) to (\ref{eqn_e}) with the previous value. We reformulate the problem as a continuous form as below.
\begin{equation}
\mathop {\min }\limits_\eta \qquad {c_{Bh}} + {c_{RAN}}\\
\end{equation}
\begin{equation}
\label{eqnr}
s.t. \quad\qquad 0 \le \eta  \le 1
\end{equation}

This continuous problem can be solved with particle swarm optimization algorithm. The algorithm finds the optimal solution by iteratively improving a candidate solution according to the objective function. A detailed description of the adapted PSO algorithm is provided below using pseudo code.

The algorithm mainly contains four steps. Firstly, the parameters used by the algorithm are initialized, including the size of the population $m$, particles' velocities $V_i$ and locations $\eta_i$, objection function values $Q_i$ and the optimal value $Q^*$, maximum iteration time $T$, and coefficient $c_1, c_2$. $v_{max}$ is the maximum limitation of a particle's velocity. Secondly, the total cost is calculated for each particle $\eta_i$, and obtain the local best solution $pbest_i$. Thirdly, it finds the global best solution of the current iteration. Finally, the particles's velocity and location are updated to start next iteration. The algorithm terminates when the maximum iteration time is reached or the error requitement is satisfied.

\begin{algorithm}[htb]
\label{alg}
\begin{algorithmic}[1]
\REQUIRE~~        
$N,M,s,F,\alpha,\beta,$

\ENSURE~~         
Optimal redundancy ratio: $\eta_{opt}$
\STATE
Initialization parameters:\\
$m\leftarrow 200$;
$V_i\leftarrow rand()*0.02-0.01$;
$\eta_i\leftarrow rand()$\\
${Q^*} \leftarrow inf$;
$Q_i \leftarrow inf$;
$pbest_i\leftarrow 0$;
$c_{total}\leftarrow inf$\\
$t \leftarrow 0 $;
$T \leftarrow 100 $;
$c_1 = 0.1$;
$c_2 = 10$;
$v_{max} = 0.0001$
\REPEAT
\STATE
$t \leftarrow  t + 1$ \\
Calculate total cost for each $\eta_i$, and obtain $pbest_i$.\\

\FOR {$i=1,2,...,m$}
\STATE $c_{total}(\eta_i) = c_{Bh}(\eta_i) + c_{RAN}(\eta_i) $

\IF {$c_{total}(\eta_i) < Q_i$}
\STATE
$Q_i \leftarrow c_{total}(\eta_i) $ \\
$pbest_i \leftarrow \eta_i$\\
\ENDIF

\ENDFOR

\STATE Find global best value of $\eta$.\\

\IF {${Q^*} > min(Q_i)$}
\STATE
${Q^*} \leftarrow min(Q_i)$  \\
$gbest \leftarrow \arg {\min _{{\eta _i}}}{Q_i}$
\ENDIF

\STATE Update particles' velocity and location.\\
$V_i \leftarrow (0.9+t/T*0.5)*V_i + c_1*rand()*(pbest_i-\eta_i)
+ c_2*rand()*(gbest-\eta_i)$
\IF {$|V_i|>v_{max}$}
\IF {$|V_i| > 0$}
\STATE $|V_i| = v_{max}$
\ELSE
\STATE $|V_i| = -v_{max}$
\ENDIF
\ENDIF

$\eta _i = \eta _i + V_i$

\UNTIL ${Q^*}$ is stable
\STATE $\eta_{opt} \leftarrow gbest$

\RETURN $ \eta_{opt} $
\end{algorithmic}
\end{algorithm}
\vspace*{-10pt}
\section{Simulation Results}
\label{sec:sim}
\vspace*{-4pt}
In this section, we present the numerical results of the change of the total transmission cost with caching redundancy ratio and verify them via Monte-Carlo simulations.

\subsection{Simulation Configuration}
We consider a cellular network where BSs are SPPP distributed in a circle of radius $r=100m$, with BS density $\lambda=2 \times 10^{-4}/m^2$. Based on the previous studies, the popularity distribution of the files follows the zipf's law, with default exponent $s=0.8$ \cite{zipf}.
The file catalog of the network contains $F=500$ different files each with normalized size of 1 byte. The default cache size of each BS is $M=50$ files. The default unit cost ratio between backhaul transmission and RAN transmission is $\mu_{BR}=4$.

We will compare the theoretic and simulation results of total transmission cost and analyze the impact of three important parameters on optimal redundancy ratio: unit transmission cost ratio of backhaul to RAN, file request pattern and cache size.

\subsection{Simulation Analysis}
\subsubsection{Impact of the unit cost ratio of backhaul to RAN transmission $\mu_{BR}$}
\begin{figure}[!t]
\centering
\includegraphics[width= 8cm]{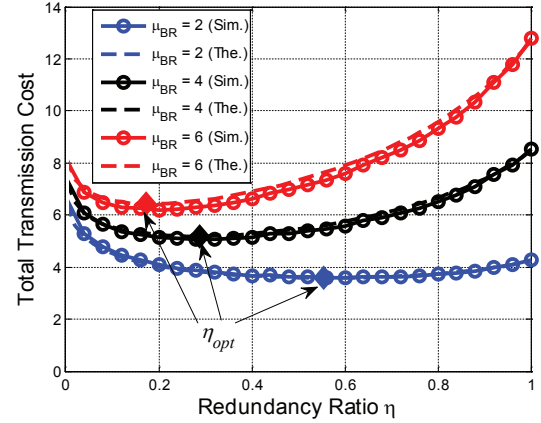}
\caption{Impact of the unit cost ratio of backhaul to RAN transmission: $\mu_{BR}$}
\label{fig_3}
\vspace*{-10pt}
\end{figure}

As shown in Fig. \ref{fig_3}, with the redundancy ratio varied from 0 to 1, the total transmission cost of the network first becomes lower, then gradually increases after passing the optimal redundancy ratio. When the unit cost ratio $\mu_{BR}$ increases, the optimal redundancy ratio decreases. This means the higher the backhaul unit transmission cost is, the more different files should be cached in BSs to reduce backhaul transmission cost since more requests are satisfied by the RAN. Additionally, compared to the caching scheme that each BS caches the same contents as others, if the redundancy ratio is optimal, the cost reduction amount is much more when the unit cost ratio $\mu_{BR}$ is higher. The total cost reduction is about 54\% when $\mu_{BR}=6$ compared to 44\% when $\mu_{BR}=4$.

\subsubsection{Impact of the file request pattern $s$}
\begin{figure}[!t]
\centering
\includegraphics[width= 8cm]{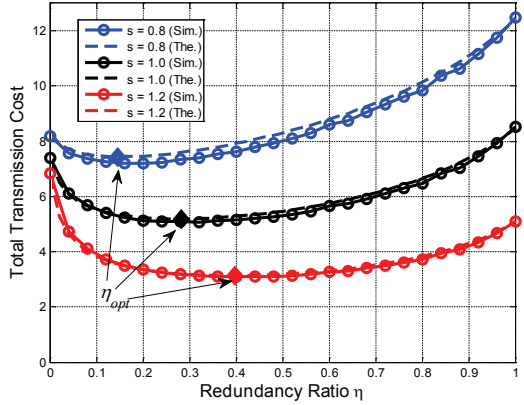}
\caption{Impact of the file request pattern: $s$}
\label{fig_4}
\vspace*{-10pt}
\end{figure}

Fig. \ref{fig_4} illustrates the influence of the exponent of file popularity distribution on the total transmission cost and optimal redundancy ratio. The greater $s$ is, the steeper file popularity distribution becomes \cite{Living}. The pattern of the transmission cost curves is similar to Fig. \ref{fig_3}. It shows that as the distribution exponent $s$ increases, the total transmission cost decreases. This is because the hit ratio of BSs' cache units increases as file distribution becomes steeper. Thus, more file requests are served locally without extra transmission. We can also observe that when $s$ increases, the optimal redundancy ratio also increases. This indicates that if the file popularity distribution is steeper, more popular files should be cached in the BSs. Under the typical file distribution when $s=0.8$, the maximum cost reduction is up to 57\% when the optimal $\eta$ is selected, compared to when $\eta=0$.

\subsubsection{Impact of the cache size $M$}
\begin{figure}[!t]
\centering
\includegraphics[width= 8cm]{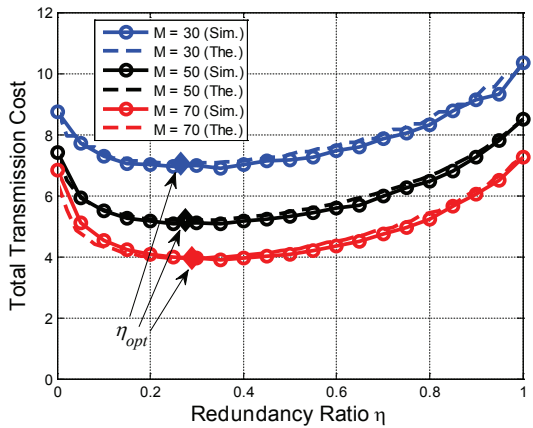}
\caption{Impact of the cache size: $M$}
\label{fig_5}
\vspace*{-10pt}
\end{figure}

Finally, Fig. \ref{fig_5} indicates that increasing cache sizes can reduce the total transmission cost under the assumption that unit transmission cost of backhaul is higher than that of RAN. This is for the reason that less backhaul transmission happens if BSs cache more files. Furthermore, the optimal redundancy ratio increases slightly as the cache size increases. This indicates that cache size does not affect the optimal caching policy too much. The optimal redundancy ratio is mainly influenced by unit cost ratio $\mu_{BR}$ and Zipf parameter $s$.

\section{Conclusion}
\label{sec:conc}
In this paper, we consider the tradeoff between caching diversity and redundancy. A novel caching scheme considering caching redundancy among BSs is proposed aiming at minimizing the total transmission cost of the network. The optimal redundancy ratio is acquired with PSO algorithm. In contrast to the traditional caching scheme that all BSs store the same most popular files, our caching scheme when choosing the optimal redundancy ratio can greatly minimize the total transmission cost. The cost reduction amount is up to 54\% when the backhaul to RAN unit cost ratio is 6. Under the typical file popularity distribution when $s=0.8$, the maximum cost reduction is up to 57\% when the redundancy ratio is optimal, compared to caching without redundancy. The optimal redundancy ratio is mainly influenced by backhaul to RAN unit cost ratio $\mu_{BR}$ and Zipf parameter $s$. Increasing cache size can lower the transmission cost but doesn't affect the optimal redundancy ratio too much.

In future, our work would include the study on caching in heterogenous network where the user demands are inhomogeneous. New caching policies will be investigated exploiting the cooperation among macro cells and small cells. In addition, the energy efficiency of caching under different scenarios is also an inspiring topic.

\section*{Acknowledgment}

This work is supported by National 973 Program under grant 2012CB316005, the National Science Foundation of China (NSFC) under grant 61372114 and 61571054, the Fundamental Research Funds for the Central Universities under grant 2014ZD03-01, the New Star in Science and Technology of Beijing Municipal Science \& Technology Commission (Beijing Nova Program: Z151100000315077), the Beijing Higher Education Young Elite Teacher Project under grant YETP0434.

\end{document}